\documentstyle[twoside,fleqn,espcrc2,amssymb,epsfig]{article}

\newcommand{\chup}{\chi U\phi}
\newcommand{\cbcex}{\langle\overline{\chi}\chi \rangle}
\newcommand{\cbc}{\overline{\chi}\chi}
\newcommand{\bt}{\beta}

\newcommand{\kp}{\kappa}

\newcommand{\chb}{\overline{\chi}}

\newcommand{\be}{\begin{equation}}
\newcommand{\ee}{\end{equation}}
\newcommand{\bea}{\begin{eqnarray}}
\newcommand{\eea}{\end{eqnarray}}
\newcommand{\half}{\frac{1}{2}}

\newcommand{\cc}{\cite}
\ifx\undefinedcommand\lesssim
\else
\fi
\newcommand{\Reop}{\mathop{\rm Re}\nolimits}

\title{%
  \mbox{}\hbox to 0pt{\vbox to 0pt{\vss
    \parbox[b]{\hsize}{\normalsize
      \begin{flushright}
        HLRZ 54/95\\
        hep-lat/9509038
      \end{flushright}
      \vspace*{2.5cm}}}\hss}%
  Tricritical point in strongly coupled U(1) gauge theory with fermions and
  scalars\footnotemark}

\author{W. Franzki\address{Institute of Theoretical Physics E, RWTH
    Aachen, D-52056 Aachen, Germany} and J. Jers\'ak$^{\rm a}$
  }

\begin{document}

\begin{abstract}
  We investigate the tricritical point in the lattice fermion--gauge--scalar
  model with U(1) gauge symmetry.  In the vicinity of this point, in the phase
  with the broken chiral symmetry, we observe the scaling behavior of the
  chiral condensate and of the masses of composite fermion and composite
  scalar, indicating the existence of an interesting continuum limit of the
  model at this point.
\end{abstract}

% typeset front matter (including abstract)
\maketitle

%This is a ugly solution, but it works
\renewcommand{\thefootnote}{\fnsymbol{footnote}}
\footnotetext{Work supported by DFG and BMBF. The computations have
  been performed on VPP500 of RWTH Aachen and CRAY-YMP of HLRZ J\"ulich.}

\section{Introduction}

The lattice 4D fermion-gauge scalar ($\chup_4$) model with U(1) gauge symmetry
has at quite strong gauge coupling, $\bt \simeq 0.64$, a tricritical point
\cc{FrJe95a,FrFr95a}.  The nature of the continuum limit taken at this point
is not known.  When this point (point E in fig.~\ref{fig:1}) is approached
from the phase with spontaneously broken global chiral symmetry and a massive
fermion $F$ (Nambu phase), the mass $am_F$ scales. The fermion $F =
\phi^\dagger\chi$ is composed of the fundamental fermion ($\chi$) and scalar
($\phi$) fields and is unconfined.  This raises the hope that a
nonperturbatively renormalizable theory with dynamical fermion mass generation
might be obtained \cc{FrJe95a}.

Motivated by this interesting possibility we have started an investigation of
the scaling behavior of various observables in the vicinity of the point E.
We find that apart from $am_F$ also the mass $am_S$ of a composite scalar
boson $S = \sum_{i=1}^{3} \phi^\dagger_{\vec{x},t} U_{(\vec{x},t), i}
\phi_{\vec{x}+\vec{i},t}$ scales in the vicinity of E.  This, as well as some
other properties \cc{FrFr95a} make the point distinctly different from the
(nonrenormalizable~\cc{AlGo95}) Nambu--Jona-Lasinio model found in the strong
coupling limit of the $\chup_4$ model~\cc{LeShr87a}.

This, as well as the experience that tricritical points give rise to the
scaling behavior different from that associated with the adjacent critical
points~\cc{LaSa84}, contributes to the hope that the point E might correspond
to a renormalizable theory.

As reported in \cc{FrJe96c}, in 2D an analogous $\chup_2$ model seems to
belong to the universality class of the 2D Gross-Neveu model and is thus
probably renormalizable.

\section{The model}
The action is
\begin{equation}
    S_{\chi U \phi} = S_\chi + S_U + S_\phi \; ,
\label{action}
\end{equation}
where
\begin{eqnarray}
  S_\chi \hspace{-2mm}&=&\hspace{-2mm} {\textstyle \half}
  \sum_x \sum_{\mu = 1}^4
  \eta_{\mu x} \chb_x \left[ U_{x,\mu} \chi_{x + \mu} -
    U_{x-\mu,\mu}^\dagger  \chi_{x-\mu} \right]\!
  \nonumber\\
  && + \, a m_0 \sum_x \chb_x \chi_x \; ,
\label{SCH}                  \\
  S_U   \hspace{-2mm}&=&\hspace{-2mm} \beta \, \sum_{\rm P} \left[ 1 - \Reop
    U_{\rm P} \right] \; ,
\label{SU}  \\
  S_\phi \hspace{-2mm}&=&\hspace{-2mm} - \kp \, \sum_x \sum_{\mu=1}^4 \left[
    \phi_x^\dagger U_{x,\mu} \phi_{x + \mu} \,+\, {\rm h.c.} \right] \; .
\label{SPH}
\end{eqnarray}
Here $U_{\rm P}$ is the plaquette product of link variables $U_{x,\mu}$ and
$\eta_{\mu x} = (-1)^{x_1 + \cdots + x_{\mu - 1}}$. The gauge field link
variables $U_{x,\mu}$ are elements of the compact gauge group U(1).  The
complex scalar field $\phi$ of charge one satisfies the constraint~$|\phi
|=1$.  The hopping parameter~$\kp$ vanishes (is infinite) when the squared
bare scalar mass is positive (negative) infinite.

The staggered fermion field $\chi$ of charge one leads to the global U(1)
chiral symmetry of the model in the chiral limit, i.e.\ when the bare fermion
mass $m_0$ vanishes.

\begin{figure*}[t]
  \begin{center}
    \leavevmode
    \epsfig{file=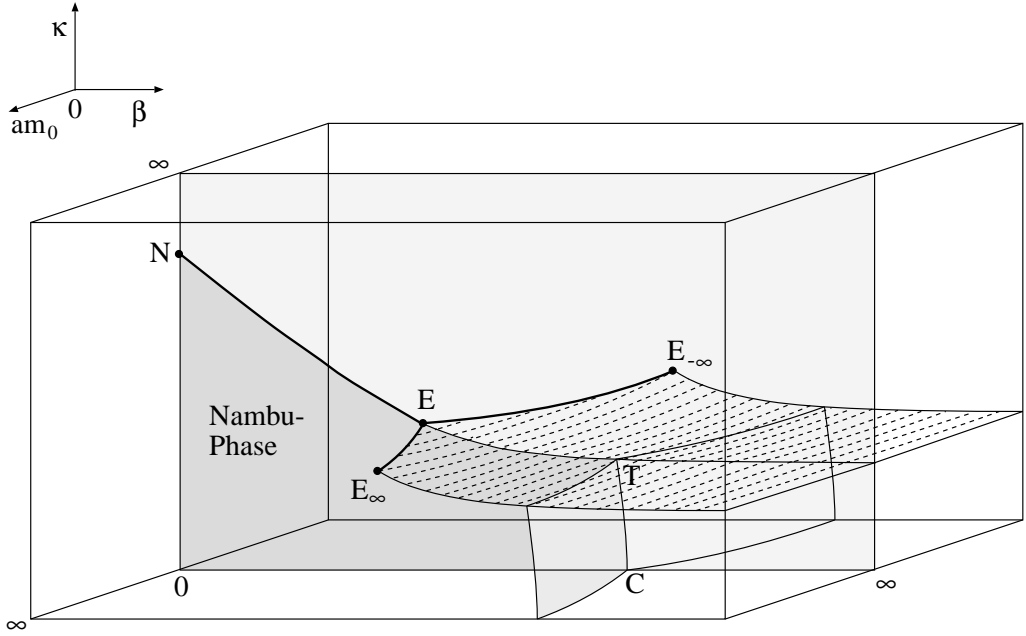,width=.85\hsize}%
    \vspace{-5mm}%
    \caption[xxx]{Schematic phase diagram of the $\chup_4$ model with U(1)
      gauge symmetry. Three critical lines, NE, EE$_\infty$ and EE$_{-\infty}$
      meet at the tricrititical point E. The NE line is a part of the boundary
      of the Nambu phase (shadowed region) at $m_0=0$, which is actually a
      sheet of 1$^{\rm st}$ order phase transitions, across which $\cbcex$
      changes sign.  The lines EE$_\infty$ and EE$_{-\infty}$ form a boundary
      of the ``wings'' (dashed sheets), which are sheets of 1$^{\rm st}$ order
      phase transitions, corresponding at large $\beta$ to the Higgs phase
      transition. The dotted part of the $m_0=0$ plane is the region of
      vanishing fermion mass, $am_F=0$. The vertical sheets containing the
      points T and C are 1$^{\rm st}$ order transitions separating the
      confinement and Coulomb phases. Line ET is a line of 1$^{\rm st}$ order
      triple points.}
    \label{fig:1}
  \end{center}
\end{figure*}
The model is meant in the chiral limit, $m_0 = 0$.  However, numerical
simulations require nonvanishing $m_0$ and thus also analytic considerations
have to include the bare mass.  We therefore show in fig.~\ref{fig:1}
schematic phase diagram of the model in the $(\bt, \kp, am_0)$ space.  It
illustrates that the tricritical point E is indeed a point in which three
critical lines meet:
\begin{itemize}
\item Line NE where $am_F$ and Goldstone boson mass $am_\pi$ simultaneously
  smoothly vanish.
\item Lines EE$_\infty$ and EE$_{-\infty}$ with vanishing
  $am_S$.
\end{itemize}
The Nambu phase below the NE line at $am_0 = 0$ is characterized by $am_F > 0$
and $am_\pi = 0$.

The limit cases are
\begin{itemize}
\item $\beta = \infty$: 4D XY model and free fermion field.
\item $\beta = 0$: Nambu-Jona-Lasinio model with bare mass.
\item $m_0=\pm\infty$: U(1) Higgs model without fermions.
\item $\kappa = 0$: compact QED with charged fermion field.
\item $\kappa = \infty$: free fermion field.
\end{itemize}

\section{Indications of scaling behavior}

As is apparent from fig.~\ref{fig:1}, an approach to the tricritical point
requires in principle tuning of three parameters. Obviously also the lattice
sizes have to be varied.  But the tricritical points have frequently rather
large domains (angles) of dominance \cc{LaSa84}.  Therefore it is reasonable
to fix $\bt = 0.64$, which is our current best estimate of the position of the
point E in the $\bt$ direction, and vary only $am_0$ and $\kp$.  Even so an
investigation of the scaling behavior is a tremendous task.  We can present
only some preliminary results obtained for $am_0 = 0.06, 0.04, 0.02$ and,
rather incomplete, also for $am_0 = 0.01$, without a proper analytic analysis
and in particular without taking into account finite size effects.

\begin{figure}[tbp]
  \vspace*{-3mm}%
  \begin{center}
    \leavevmode
    \hspace*{-8mm}%
    \epsfig{file=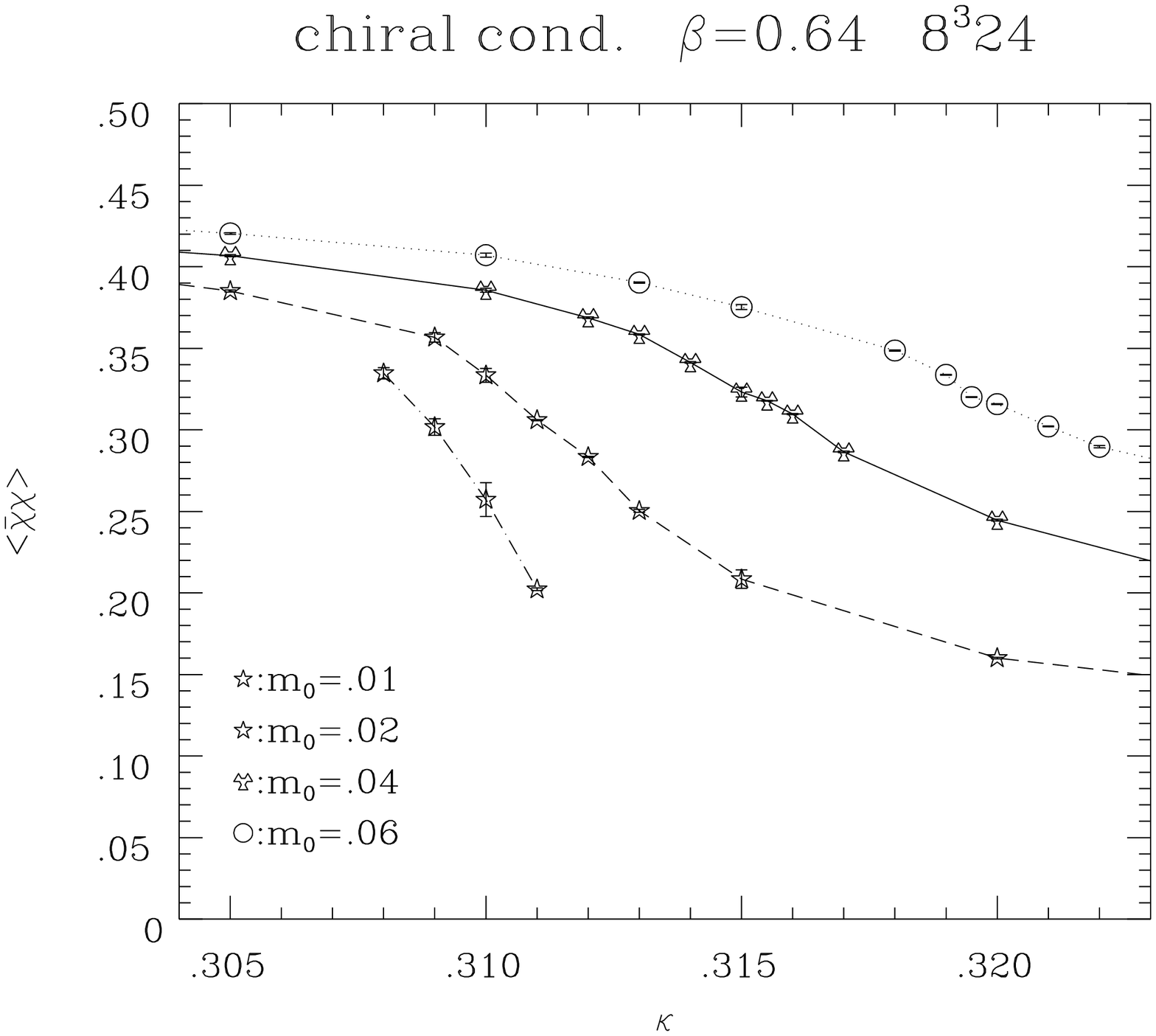,width=7cm,angle=90}\hspace*{-18mm}%
    \vspace*{-10mm}%
    \caption{Chiral condensate $\cbcex$.}
    \label{fig:2}
  \end{center}
\end{figure}
\begin{figure}[tbp]
  \vspace*{-3mm}%
  \begin{center}
    \leavevmode
    \hspace*{-8mm}%
    \epsfig{file=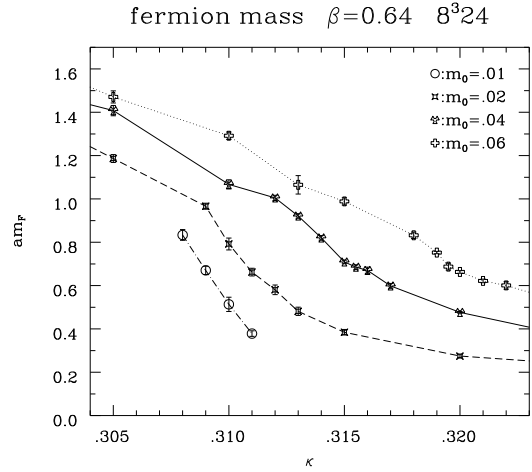,width=7cm,angle=90}\hspace*{-18mm}%
    \vspace*{-10mm}%
    \caption{Fermion mass $am_F$.}
    \label{fig:3}
  \end{center}
\end{figure}
In fig.~\ref{fig:2} we show the chiral condensate $\cbcex$.  An onset of the
genuine chiral phase transition is indicated with decreasing $am_0$.  The
fermion mass $am_F$ (fig.~\ref{fig:3}) behaves similarly.  The transition
appears to be smooth, i.e. of 2$^{\rm nd}$ order, close to $\kappa\simeq
0.305$.

\begin{figure}[tbp]
  \vspace*{-3mm}
  \begin{center}
    \leavevmode
    \hspace*{-8mm}%
    \epsfig{file=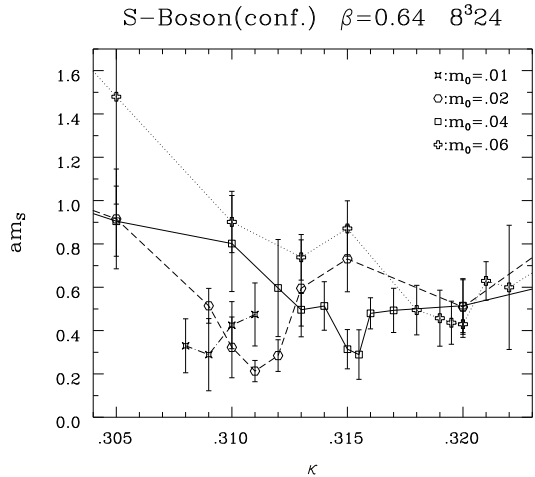,width=7cm,angle=90}\hspace*{-18mm}%
    \vspace*{-10mm}%
    \caption{Scalar boson mass $am_S$.}
    \label{fig:4}
  \end{center}
\end{figure}
In fig.~\ref{fig:4} we show the scalar boson mass $am_S$.  The position of the
minimum for each $am_0$ corresponds roughly to the crossection of a
continuation to smaller $\beta$ of the nearly horizontal ``wing'' with the
$\bt = 0.64$ plane.  As $am_0$ decreases, the minimum seems to approach the
$\kp$ value at which the chiral phase transition at $am_0 = 0$ is expected on
the basis of figs. 2 and 3.  Thus we are indeed in a close vicinity of the
tricritical point.

\begin{figure}[tbp]
  \vspace*{-3mm}%
  \begin{center}
    \leavevmode
    \hspace*{-8mm}%
    \epsfig{file=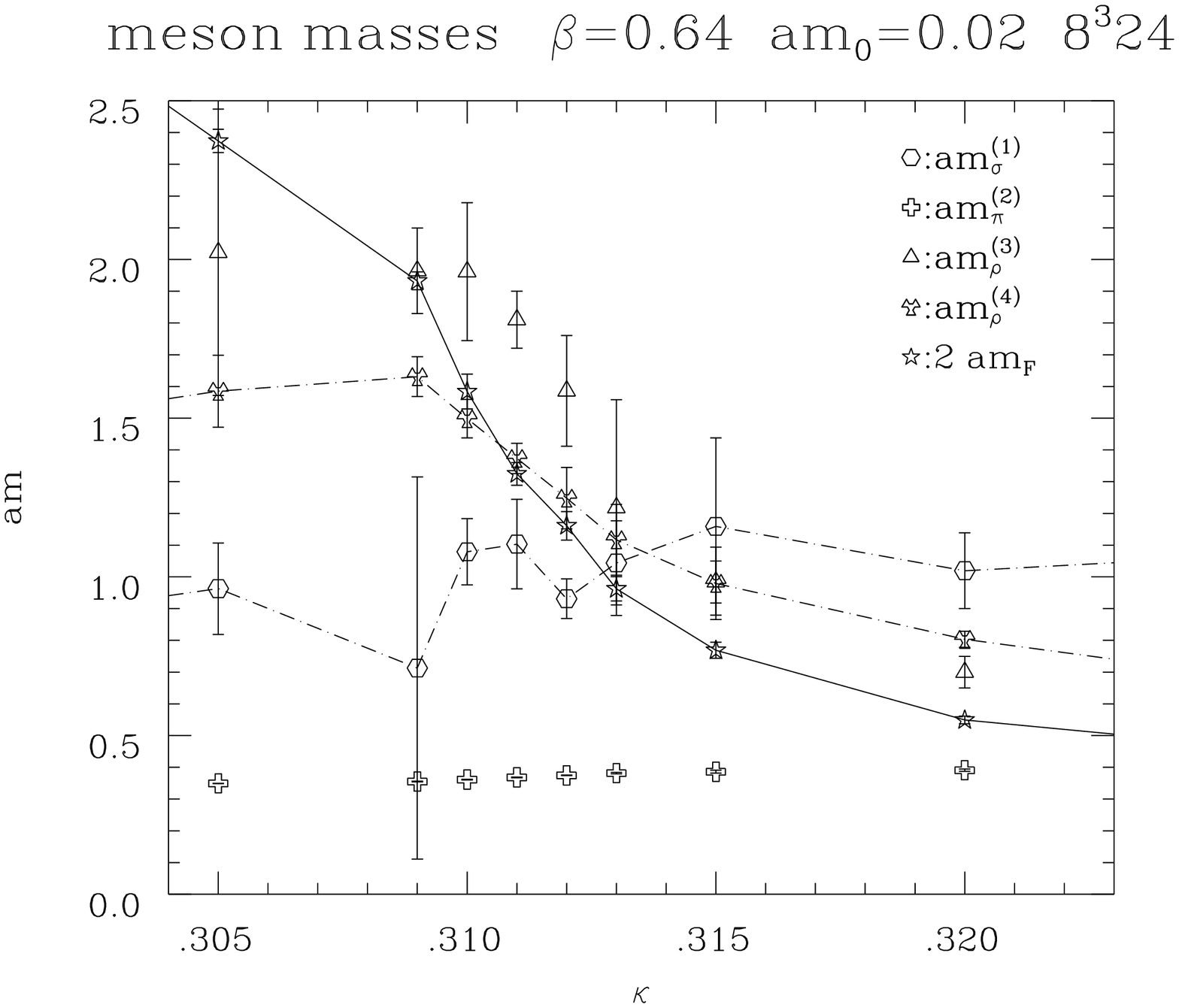,width=7cm,angle=90}\hspace*{-18mm}%
    \vspace*{-10mm}%
    \caption{Masses of mesons ($\cbc$ composite states). These mesons are
      defined in \protect\cc{FrJe95a}.}
    \label{fig:5}
  \end{center}
\end{figure}
Fig.~\ref{fig:5} contains very preliminary results for several $\cbc$ (meson)
masses.  For comparison we show also data for $2am_F$.  The Goldstone boson
mass $am_\pi$ is small around the transition as expected.  The fact that the
$\rho$- and $\sigma$- masses cross the $2am_F$ line in the vicinity of the
phase transition indicates that in the Nambu phase these mesons can be
interpreted as bound states and that their masses scale similarly to $am_F$
and thus may exist also in the continuum limit.  This is a hint that the
spectrum of the underlying continuum theory might be quite complex.

Our present results suggest that it is worthwhile to continue the
investigation of the tricritical point E. For this purpose the full scaling
theory of tricritical points~\cc{LaSa84} and a finite size scaling theory in
their vicinity will have to be applied. Some analytic insight into the
properties of the model in the relevant coupling region is highly desirable.

\bibliographystyle{wunsnot}
% \bibliography{jourabbr,our-papers,gauge,yukawa,referen}

% end_of_bibliography
\end{document}